\documentstyle[seceq,preprint]{jpsj}
\def\theequation{\arabic{section}.\arabic{equation}}

\def\img{\mbox{i}}
\def\ostr{O_{\mbox{str}}}
\def\H{{\cal H}}
\def\v#1{\mbox{\boldmath$#1$}}

\def\runtitle{DMRG Study of Random Dimerized AF Heisenberg Chains
}

\def\runauthor{Kazuo {\sc Hida}}

\title
{Density Matrix Renormalization Group Study\\ of Random Dimerized Antiferromagnetic Heisenberg Chains
}

\author
{Kazuo {\sc Hida}
\footnote{e-mail: hida@riron.ged.saitama-u.ac.jp}}

\inst
{
Department of Physics, Faculty of Science,\\ Saitama University, Urawa, Saitama 338
}

\recdate
{
May 14, 1997}

\abst
{
The effect of dimerization on the random antiferomagnetic Heisenberg chain with spin 1/2 is studied by the density matrix renormalization group method. The ground state energy, the energy gap distribution and the string order parameter are calculated. Using the finite size scaling analysis, the dimerization dependence of the these quantities are obtained. The ground state energy gain due to dimerization behaves as $u^a$ with $a > 2$ where $u$ denotes the degree of dimerization, suggesting the absence of spin-Peierls instability.  It is explicitly shown that the string long range order survives even in the presence of randomness. The string order behaves as $u^{2\beta}$ with $\beta \sim 0.37$ in agreement with the recent prediction of real space renormalization group theory ($\beta =(3-\sqrt{5})/2 \simeq 0.382$). The physical picture of this behavior in this model is also discussed. }

\kword
{
density matrix renormalization group, random dimerized antiferromagnetic Heisenberg chain, renormalization group theory
}

\begin{document}

\maketitle

\section{Introduction}
In the recent studies of quantum many body problem, the ground state properties of the random quantum spin systems have been attracting a renewed interest owing to the occurence of exotic phases resulting from the interplay of quantum fluctuation and randomness. The real space renormalization group (RSRG) analysis of the $S=1/2$ random antiferromagnetic Heisenberg chain (RAFHC)\cite{bl1,dsm1,dsm2,ds1} has shown that the ground state of this model is the random singlet (RS) state in which the spins form singlets randomly with distant partners. This has been confirmed numerically by means of the density matrix renormalization group calculation.\cite{kh1}

On the other hand, the effect of randomness on the spin gap state of one-dimensional quantum spin systems has been extensively studied theoretically and experimentally\cite{hy1,hy2,f1,f2,sf1,nf1,moto1,ii1,mik1,fm1,st1,ren1,hase1} in the relation with the doping effect in spin ladders and spin-Peierls systems. Hyman {\it et al.}\cite{hy1} have applied the RSRG method to the spin-1/2 random dimerized antiferromagnetic Heisenberg chain and have shown that the dimerization is a relevant perturbation to the random singlet phase. They also argued that the ground state of this model is the random dimer (RD) phase in which the string long range order survives even in the presence of randomness.

It is the purpose of the present work to investigate the ground state of the spin-1/2 random dimerized antiferromagnetic Heisenberg chain numerically using the density matrix renormalization group method.\cite{kh1,wh1,wh2} In the next section, we introduce the model Hamiltonian. The result of the numerical calculation is presented in \S 3. The last section is devoted to summary and discussion.

\section{Model Hamiltonian}
The Hamiltionan of the spin 1/2 random dimerized Heisenberg chain is given by,
\begin{equation}
\label{eq:ham}
\H = \sum_{i=1}^{N} 2J_i\v{S}_{i}\v{S}_{i+1},\ \ \mid \v{S}_{i}\mid = 1/2,
\end{equation}
with 
\begin{equation}
J_i=J(1+(-1)^{i-1}u + \delta_i),
\end{equation}
where $\delta_i$ takes random values. Here we assume the distribution

\begin{equation}
P(\delta_i) = \left\{\begin{array}{ll}
1/W & \mbox{for} -W/2 < \delta_i < W/2, \\
0 &  \mbox{otherwise}.
\end{array}\right.
\end{equation}

In the following, we define the energy unit by setting $J=1$ and assume  $u > 0$ without loss of generality. The numerical calculation is done for $W=1$. The bond which connects the $i$-th site and $i+1$-th site  is called the $i$-th bond. The bonds (sites) with even (odd) $i$ are called even (odd) bonds (sites). 

The ground state of the undimerized counterpart of this model ($u=0$) is the RS phase as verified by the RSRG method for the distribution function of the effective bond strength.\cite{dsm1,dsm2,ds1} In this phase, the effective energy scale $\Omega$ and spatial scale $L$ of a singlet pair is related as $\Gamma \equiv \ln\Omega \propto -\sqrt L$. This implies that the spins form a totally singlet state in a segment of average length $L$ at the energy scale $\Omega$. The size of each singlet segment grows as the energy scale is lowered and finally all spins form a totally singlet RS state. This calculation is extended to the dimerizied case by Hyman {\it et al}.\cite{hy1} In this case, the similar description holds down to the energy scale $\Gamma \sim 1/u$. At this energy scale, the average spatial scale of each singlet pair is $1/u^2$. As the decimation further proceeds, however, the odd bonds are renormalized to zero and the ground state is decribed as an assembly of RS segments whose average length is proportional to $1/u^2$. This phase is called the RD phase. The spin-spin correlation length is finite and proportional to $1/u^2$ but the energy gap is distributed down to zero. Therefore this phase is a Griffith phase. Hyman {\it et al}.\cite{hy1} also argued that the string order defined for the regular bond-alternating chain\cite{khreg1} remains finite in the presence of randomness. They claimed that the string order is proportional to $u^{2\beta}$ with $\beta=2$\cite{hy1} for small $u$. Quite recently, however, Hyman and Yang\cite{hy2} have given a different estimation as $\beta=(3-\sqrt{5})/2$.

\section{Numerical Results}

\subsection{Average Energy Gap}

In Fig. \ref{fig:gap0}, the average of the logarithm of the energy gap $\overline{\ln \Delta}$ is plotted against $\sqrt N$ for $u = 0$ and $N \leq 60$ using approximately 250 samples. Here the bar denotes the sample average. For all these samples, the number $m$ of the states kept in each DMRG step is 100 and only infinite size iterations are carried out. For the numerical caluculation, the samples are grouped into several subgroups. The error bars are estimated from the fluctuation among these subgruoups. For $u=0$,  the average $\overline{\ln \Delta}$ is well fitted by $\overline{\ln \Delta} \simeq 1.77-0.8\sqrt N$ as shown in Fig. \ref{fig:gap0}. This also implies that the average $\overline{\ln \Delta}$ is calculated accurately enough with $m=100$. Considering that the gap tends to increase with $u$, the accuracy for the cases $u>0$ is expected to be even better than the undimerized case $u=0$. Therefore we take $m=100$ for the calculation of  $\overline{\ln \Delta}$ for finite $u$. Hyman {\it et al}.\cite{hy1} have shown that $\overline{\ln \Delta} $ should behave as $1/u$ for finite $u$ and correlation length is proportional to $1/u^2$. Therefore we may assume the following finite size scaling (FSS) formula.
\begin{equation}
\label{eq:gsc}
\overline{\ln (\Delta/\Delta_0)} = \frac{1}{u}f(u\sqrt N) \ \ \mbox{with}\ \ln\Delta_0=1.77.
\end{equation}

The FSS plot is shown in Fig. \ref{fig:gap}. The data collapse onto a single curve fairly well confirming the conclusion of Hyman {\it et al}.\cite{hy1} We therefore analyse our numerical data assuming the FSS plot with the argument $uN^{1/2}$ in the following.

\subsection{Ground State Energy}

The average ground state energy per site $\overline{E_G}/N$ is also calculated for $N \leq 60$ using approximately 1200 samples for various values of $u$. The average magnetic energy gain per spin $\overline{\Delta E_G}/N$ due to dimerization is defined by $(\overline{E_G(u)}-\overline{E_G(u=0)})/N$. To check the accuracy, we have carried out the calculation with $m=100$ and $60$ for 50 samples. The $m$-dependence was negligible compared to $\Delta E_G$ even for $u=0.01$. Therefore, we have taken $m=60$ for the calculation of the ground state energy. Assuming the FSS formula,

\begin{equation}
\label{eq:ensc}
\overline{\Delta E_G}/N = N^{-a/2} g(u\sqrt N), 
\end{equation}
we find good fit to a single curve for $a \simeq 2.46 \pm 0.10$. Fig. \ref{fig:eg}(a) shows a FSS plot with $a=2.46$. This indicates that $\overline{\Delta E_G}/N$  behaves as $u^{2.46 \pm 0.10}$ for small $u$ in contrast to the regular dimerized chain\cite{oka} in which  $\Delta E_G/N \propto u^{4/3}$. Especially, it is clear that the FSS plot with $a=2$ does not collapse onto a single curve  as shown in Fig. \ref{fig:eg}(b) which ensures that $a$ is definitely larger than 2. This implies that the spin-Peierls {\it instablity} for infinitesimal $u$ is unlikely in the random Heisenberg chain because the cost of the lattice deformaton energy is proportional to $u^2$. 

\subsection{String Order Parameter}

The string order parameter $\overline{\ostr}$ is defined for this system by extending the definition for the regular system as follows,\cite{hy1,khreg1}

\begin{equation}
\label{eq:ost1}
\overline{\ostr} = \lim_{l, N \rightarrow \infty} \overline{\ostr(l;N)},
\end{equation}
where $\ostr (l;N)$ is the string correlation function in the chain of length $N$ which is defined only for odd $l$ as, 
\begin{equation}
\label{eq:ost2}
\ostr (l;N) =  <\exp \left\{\img\pi \sum_{k=2i+1}^{2i+l+1} S_{k}^z \right\} >_N,
\end{equation}
where $<...>_N$ denotes the ground state average for each sample of length $N$. Although Hyman {\it et al}.\cite{hy1} defined the string order parameter in somewhat different way, both definitions are equivalent in the thermodynamic limit. We have chosen this expression because the numerical error bars estimated as described in \S 3.1 are less pronounced. The sites $k=2i+1$ and $2i+l+1$ are fixed as distant as possible from the boundary for each $l$ to reduce the boundary effect. The average is taken over 400 samples with $N \leq 60$ keeping 100 states in each step. Comparing the cases with $m=60$ and $100$, we have found only slight difference. Therefore we use the data with $m=100$ in the following analysis.

Figure \ref{fig:str1} shows $\ostr (l; N)$ as a function of $l$ for relatively large values of $u (=0.1, 0.2$ and $0.3$). It should be noted that $\ostr(N-1;N) \equiv 1$ because $\sum_{k=1}^N S_{k}^z=0$ in the ground state. For these values of $u$, it is clearly seen that the string long range correlation remains finite. To determine the precise $u$-dependence of $\overline{\ostr}$ for smaller values of $u$, we resort to the FSS analysis. The string order parameter for the finite system $\overline{\ostr}(N)$ is defined by $\overline{\ostr}(N/2-1; N )$. We assume the FSS formula for $\overline{\ostr}(N)$ as,

\begin{equation}
\label{eq:stsc}
\overline{\ostr}(N) = N^{-\beta}h(u\sqrt N). 
\end{equation}
For $ \beta = 0.37 \pm 0.03$, the data fall on a single curve within the error bars. In Fig. \ref{fig:str2}(a), the FSS plot is shown with $\beta=0.37$. This indicates that $\overline{\ostr}$ is proportional to $u^{0.74\pm0.06}$ in the thermodymamic limit. This is in good agreement with the new estimation of $\beta=(3-\sqrt{5})/2$ in ref. \citen{hy2}. Figure \ref{fig:str2}(b) shows the FSS plot with $\beta=2$ as predicted by Hyman {\it et al}.\cite{hy1} at first. All curves do not collapse at all. Thus the possiblity $\beta=2$ is totally excluded within our numerical data. 

This exponent $\beta$ is related to the exponent $\eta$ of the size dependence of $\overline{\ostr}(N)$ for $u=0$ in the following way. The RD ground state can be regarded as an assembly of almost decoupled RS segments. The left ends of each segment is the odd sites and right ends, the even sites. (It is assumed that the sites are arranged in increasing order of $i$ from left to right.) The string correlation between the spins on both ends of a segment is perfect because $\sum_i S^z_i=0$ within each RS segment. Therefore this contribution is proportional to the square of the density of these boundary spins. Hyman {\it et al}.\cite{hy1} estimated $\beta = 2$ counting the contribution only from these bounadry spins. However, the finite string correlation can remain between the boundary spins and those inside the RS segments. This contribution to the string correlation between the $2i+1$-th spin and $2j$-th spin for large $j-i>>1$ is estimated as follows. 

\begin{eqnarray}
\label{eq:ostd1}
\overline{\ostr\left(l; N \right)}&=& \overline{<\exp \left\{\img\pi\left(\sum_{k \geq 2i+1; k \in I} S_{k}^z + \sum_{k \in I+1,..,J-1} S_{k}^z \right. \right.}\nonumber \\
&&\overline{+\left.\left. \sum_{k \leq 2j; k \in J} S_{k}^z \right) \right\} >_N}.
\end{eqnarray}
where $2j=2i+l+1$ and the capital indices distinguish the RS segments. It is assumed that the $2i+1$-th and $2j$-th spins belong to the different RS segments denoted by $I$ and $J$, respectively. This expression is approximated as
\begin{eqnarray}
\label{eq:ostd2}
\overline{\ostr\left(l; N \right)} &\simeq& \overline{<\exp \left\{\img\pi \sum_{k \geq 2i+1; k \in I} S_{k}^z \right\} >_N}  \nonumber \\
&\times& \overline{<\exp \left\{\img\pi \sum_{k \in I+1,..,J-1} S_{k}^z \right\} >_N}  \nonumber \\
&\times& \overline{<\exp \left\{\img\pi \sum_{k \leq 2j; k \in J} S_{k}^z \right\} >_N}.
\end{eqnarray}
because the correlation between the segments is weak. The second average is estimated as 

\begin{equation}
\label{eq:ostd3}
<\exp \left\{\img\pi \sum_{k \in I+1,..,J-1} S_{k}^z \right\} >_N \simeq 1,
\end{equation}
because these spins form an almost complete singlet state in each segment. If we assume that the string correlation between the boundary spin and the inner spins in a single RS segment decay by the power law as 
\begin{equation}
\overline{<\exp \left\{\img\pi \sum_{k \geq 2i+1; k \in I} S_{k}^z  \right\} >_N} \sim d_{2i}^{-\eta},
\end{equation}
\begin{equation}
\overline{<\exp \left\{\img\pi \sum_{k \leq 2j; k \in J}  S_{k}^z \right\} >_N} \sim d_{2j-1}^{-\eta},
\end{equation}
 where $d_{2i}$ ($d_{2j-1}$) is the distance between the $2i$-th ($2j-1$-th) spin and the right (left) boundary spin of the $I$-th ($J$-th) segment. Here the bars should be understood as the random average  within each segment. Thus we find
\begin{equation}
\label{eq:ostd4}
\overline{\ostr\left(l; N \right)} \sim   \overline{d_{2i}^{-\eta} d_{2j-1}^{-\eta}}.
\end{equation}
The bar on the rhs represents the average over the whole sample. The averge size of the RS segment is proportional to $1/u^2$. Therefore the average string order $\overline{\ostr}$ is proportional to $u^{-4\eta}$ which yield $\beta=2\eta$.

\begin{figure}
\caption{The log-log plot of the string correlation $\overline{\ostr}(N)$ in the undimerized random Heisenberg chain($u=0$) plotted against $N$. The solid line is the fit by the power law $\overline{\ostr}(N) \sim N^{-2\eta}$ with $\eta = 0.2$. }
\label{fig:str0}
\end{figure}
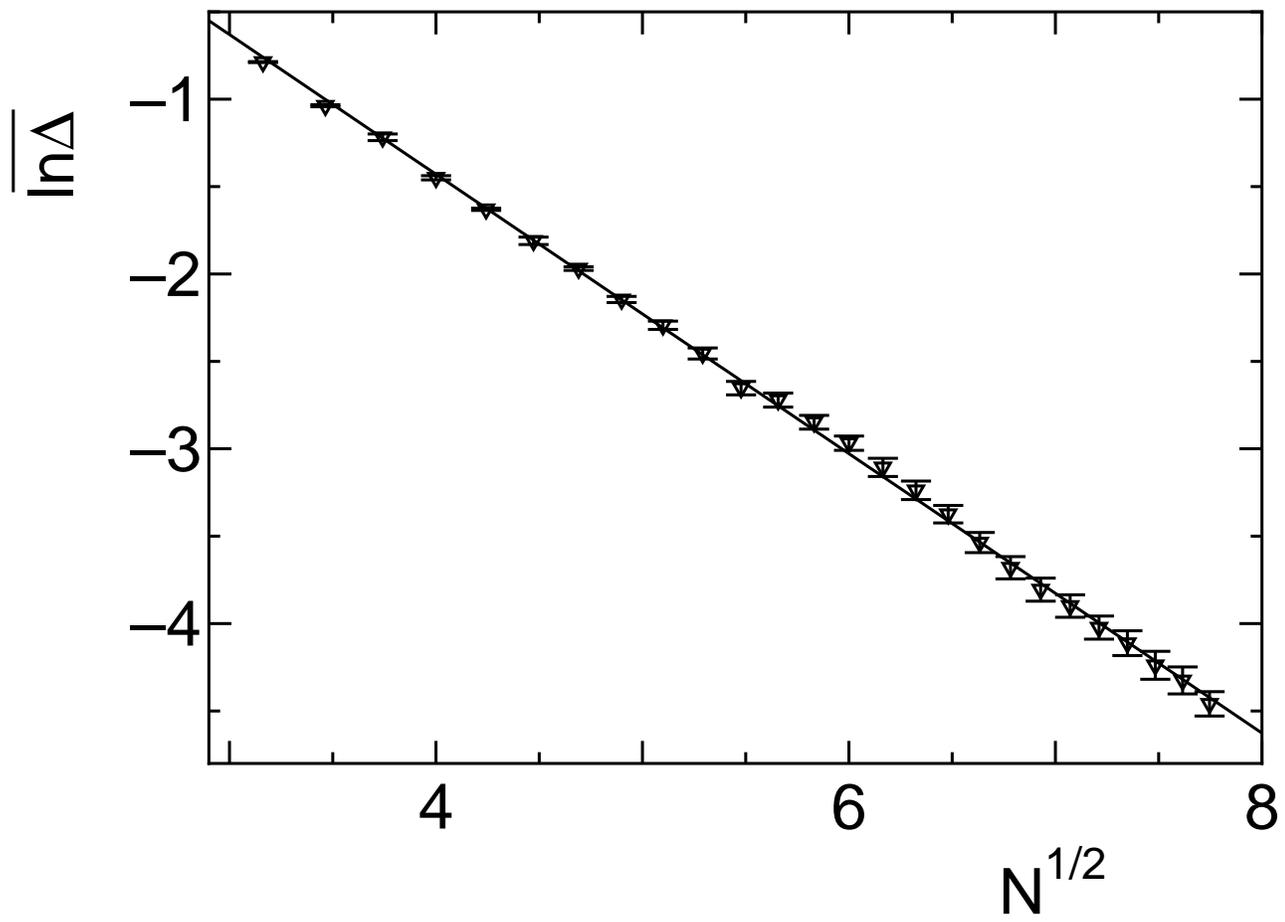
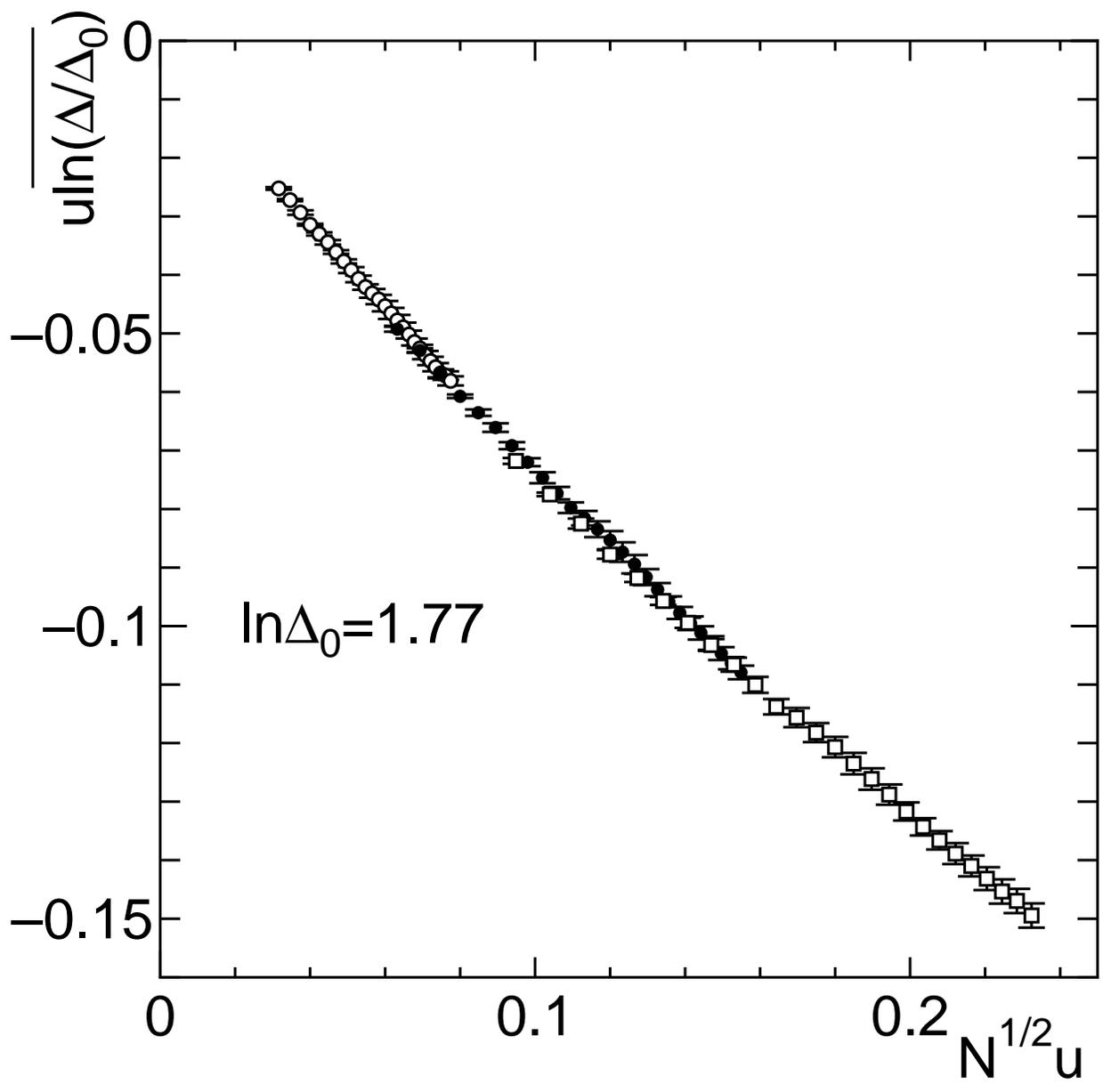
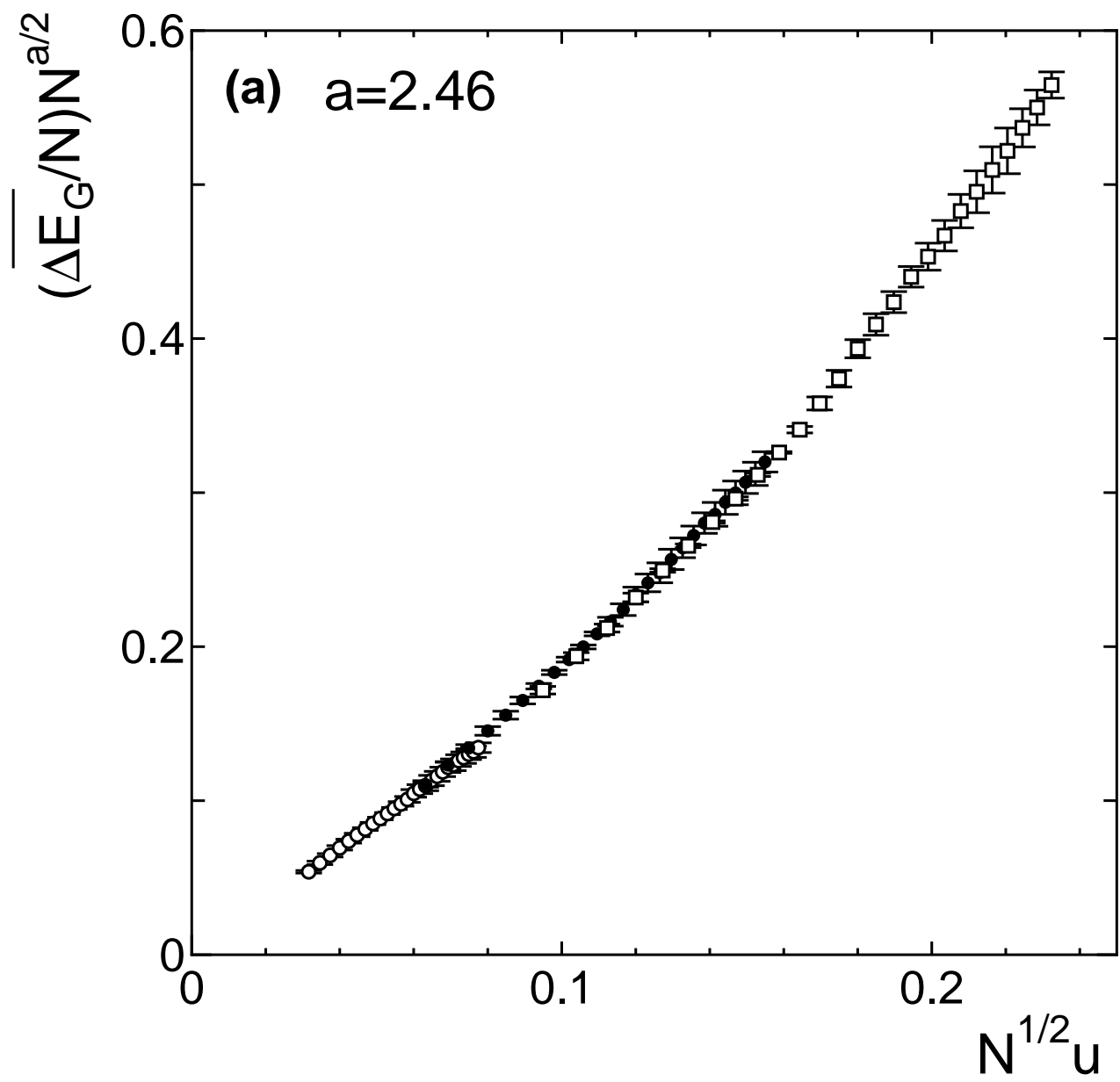
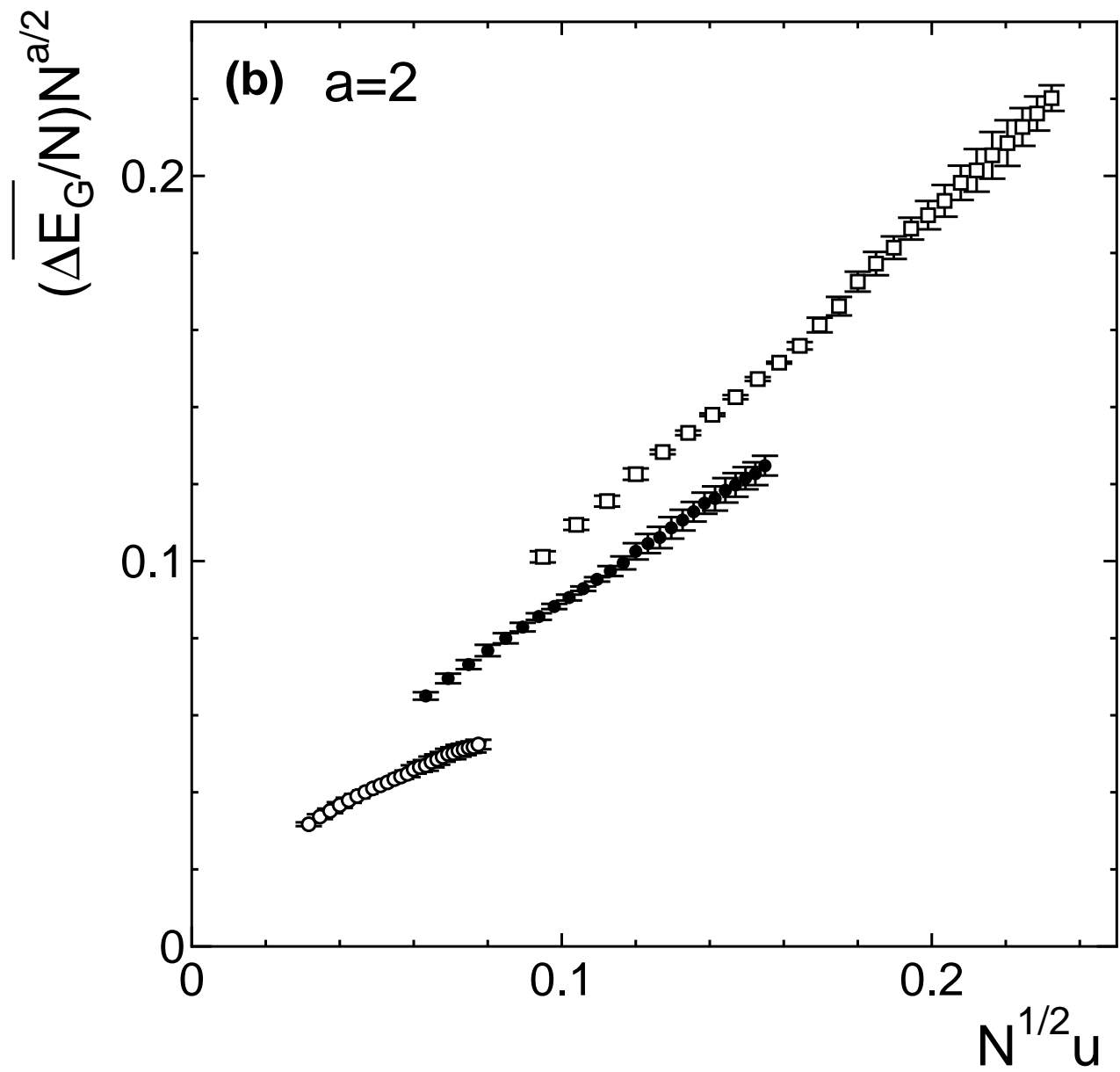
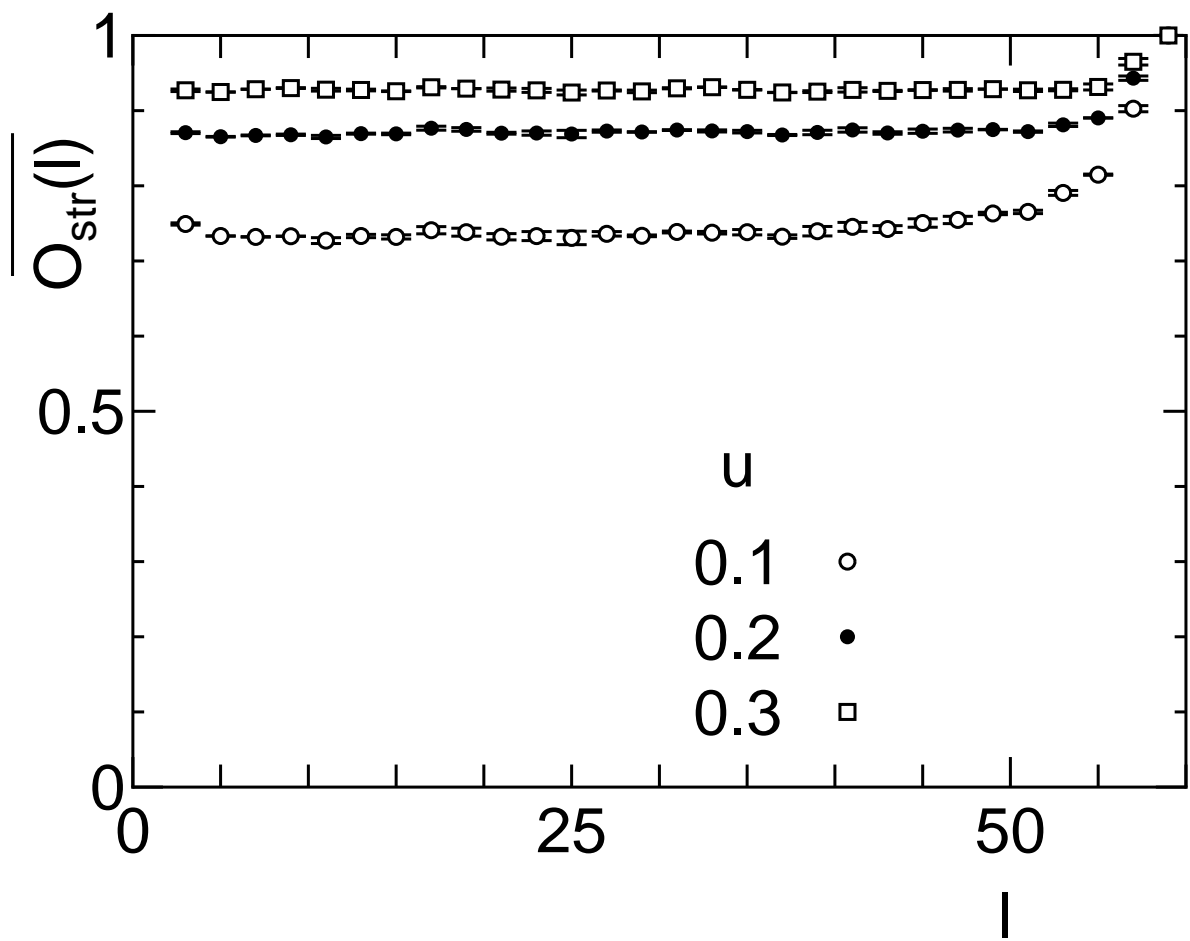
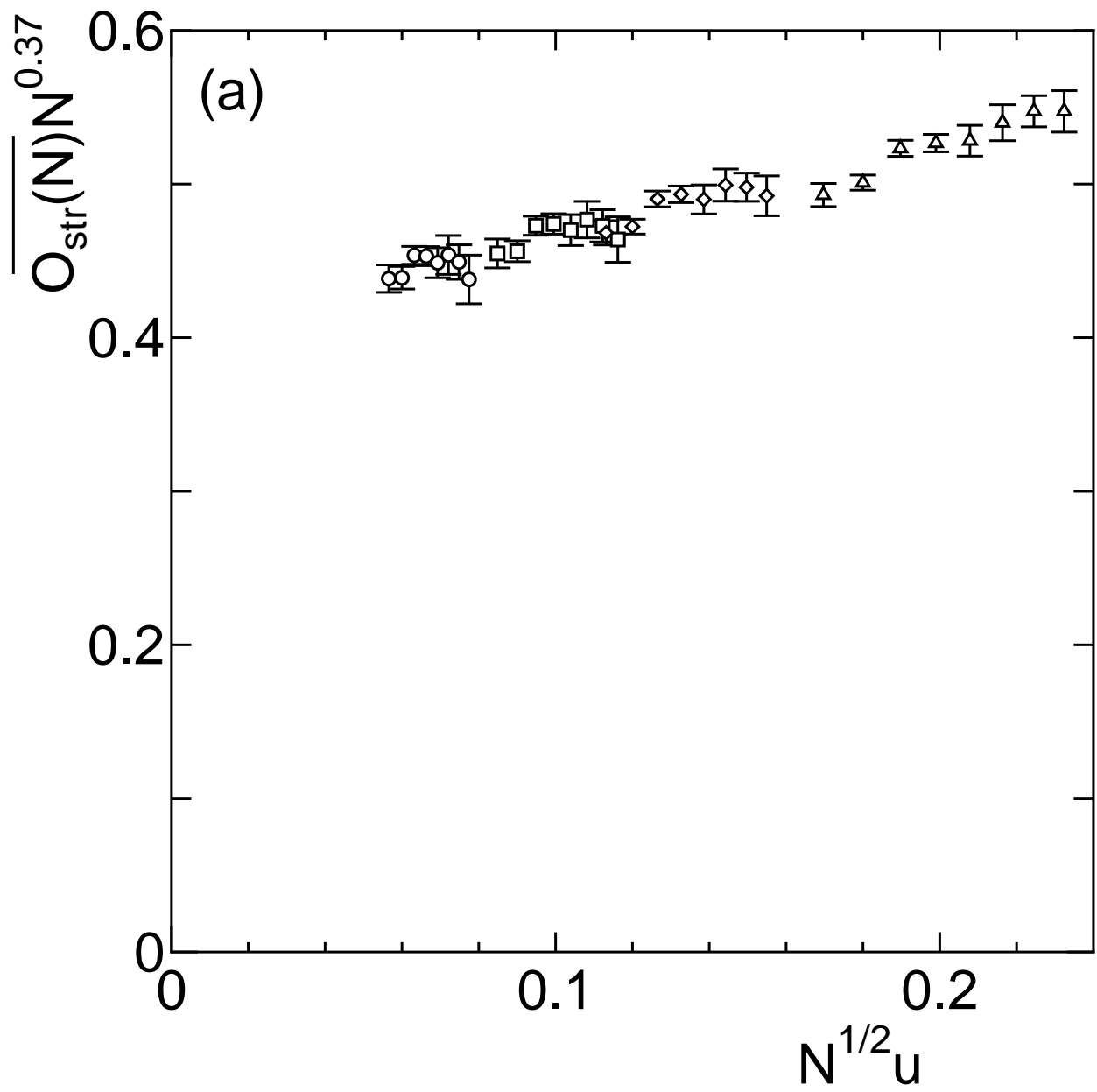
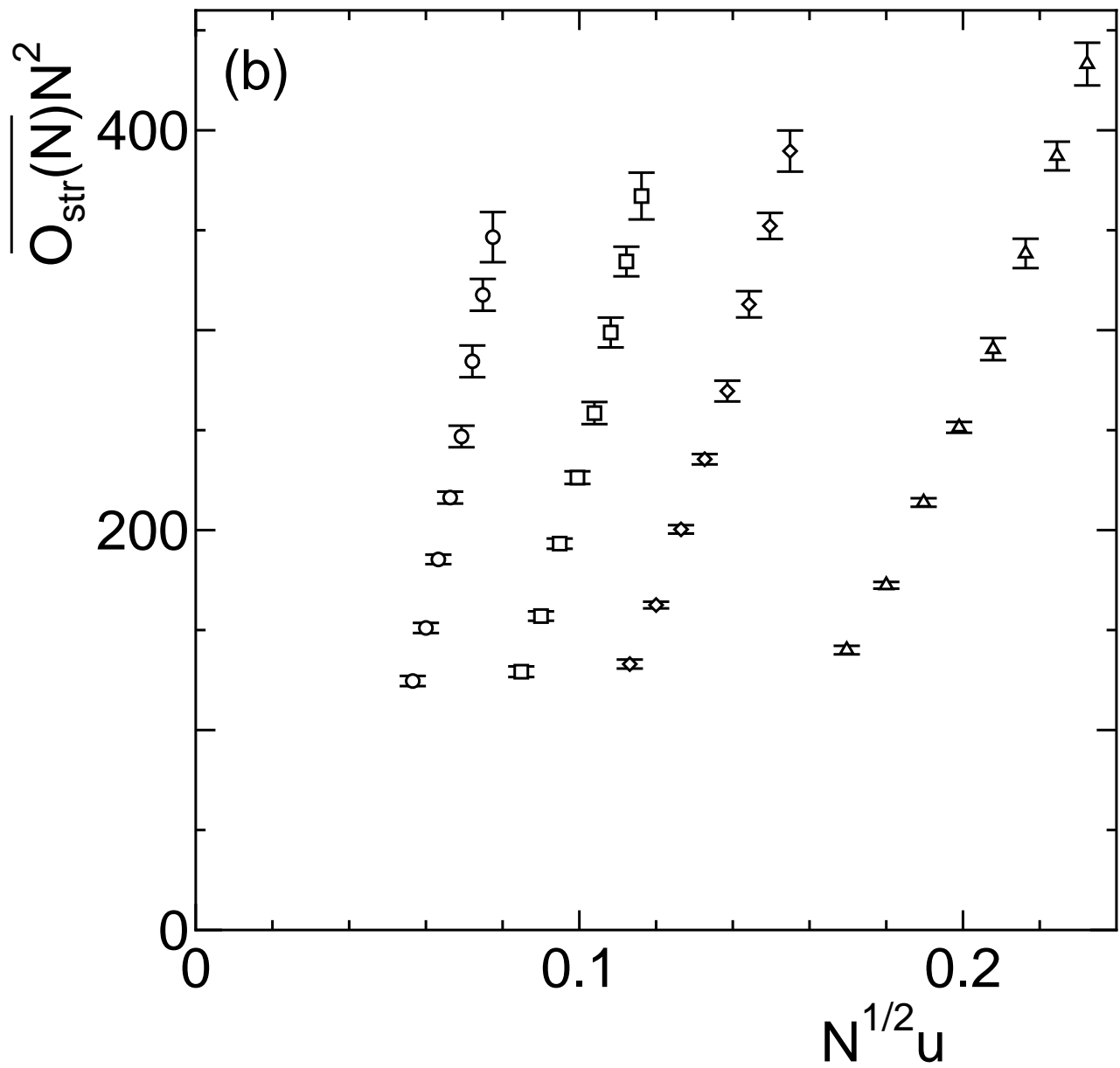
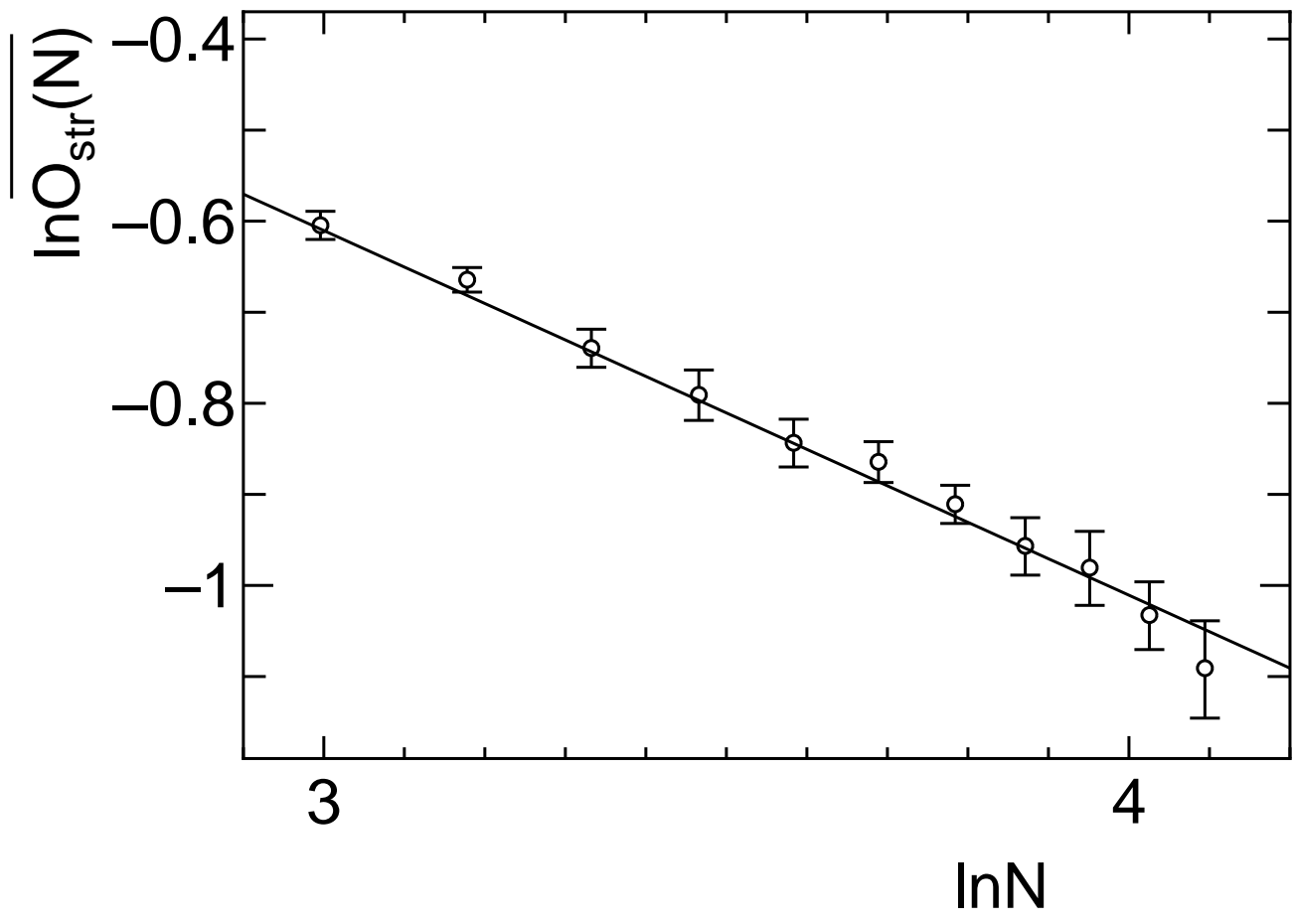

In order to estimate the value of $\eta$, we have measured the string correlation in the finite length undimerized random Heisenberg model. If the above picture holds, $\overline{\ostr}(N/2-1;N)$ can be estimated between the $N/4+1$-th site and $3N/4$-th site as follows 
\begin{eqnarray}
\label{eq:ostd5}
\lefteqn{\overline{\ostr}(N)= \overline{\ostr}(N/2-1,N)} \nonumber \\
&=& \overline{<\exp \left\{\img\pi \sum_{k=N/4+1}^{3N/4} S_{k}^z \right\} >_N} \nonumber \\&=& \overline{<\exp \left\{\img\pi \sum_{k=1}^{N} S_{k}^z - \sum_{k=1}^{N/4} S_{k}^z - \sum_{k=3N/4+1}^{N} S_{k}^z \right\} >_N} \nonumber \\
&=& \overline{<\exp \left\{\img\pi \sum_{k=1}^{N/4} S_{k}^z \right\} >_N^* <\exp \left\{\img\pi  \sum_{k=3N/4+1}^{N} S_{k}^z \right\} >_N^*} \nonumber \\
 &\sim& (N/4)^{-\eta} (N/4)^{-\eta}.
\end{eqnarray}
Here it is assumed that $\sum_{k=1}^{N} S_{k}^z=0$ because the ground state is singlet. Thus we find,
\begin{equation}
\label{eq:ostd6}
\overline{\ostr(N)} \propto N^{-2\eta}.
\end{equation}
From the log-log plot in Fig. \ref{fig:str0}, we find $2\eta \sim 0.4 \pm 0.05$ which is close to the estimation $\beta(=2\eta )\sim 0.37$ determined from the $u$-dependence of $\ostr$.

\section{Summary and Discussion}

The ground state properties of the spin-1/2 random dimerized antiferromagnetic Heisenberg chain is studied using the DMRG method. It is verified that the average of the logarithm of the energy gap is inversely proportional to the degree of dimerization in agreement with the real space renormalization group theory. 

The energy gain per spin is found to be proportional to $u^a$ with $a > 2$. Therefore the spin-Peierls instability does not take place in the present model. This mechanism of the absence of spin-Peierls instability should be distinguished from the topological argument by Hyman {\it et al}.\cite{hy1} which excludes the possibility of spontanuously dimerized ground state in purely one-dimensional random spin system. In the spin-Peierls system, the phonons should be treated  three dimensionally and the confinement of domain walls does not take place. It should be also emphasized that this argument concerns the spin-Peierls {\it instability} in the limit of weak spin-phonon coupling and does not exclude the possiblity of the spin-Peierls state for strong enough spin-phonon coupling. Actually, for weak disorder, it is reasonable to expect that energy gain $\Delta E/N$ behaves as $u^{4/3}$ as far as $u >> W$. Therefore, the spin-Peierls state would be stable against disorder, if the lattice distortion $u_0$ in the absence of randomness satisfies this condition $u_0 >> W$.  

The string order behaves as $u^{2\beta}$ with $\beta \sim 0.37$ in agreement with the prediction of Hyman and Yang.\cite{hy2} It is suggested that the discrepancy from the earlier estimation by Hyman {\it et al}.\cite{hy1} is due to the string correlation between the inner spins in different RS segment which was not taken into account in ref. \citen{hy1}. This picture is consistent with the behavior of the string correlation in the undimerized random Heisenberg chain. 

In this work, we have fixed the strength of randomness $W=1$. Nevertheless, we expect our conclusion is universal at least qualitatively as long as $W >> u$. It should be noted, however, the physics can be totally different for other kind of randomness such as the case of topological disorder studied in refs. \citen{fm1} and \citen{st1}.

The numerical calculations have been performed using the FACOM VPP500 at the Supercomputer Center, Institute for Solid State Physics, University of Tokyo.  This work is supported by Grant-in-Aid for Scientific Research from the Ministry of Education, Science and Culture.

\begin{figure}
\caption{The system size dependence of $\overline{\ln \Delta}$ for $u=0$.}
\label{fig:gap0}
\end{figure}
\begin{figure}
\caption{The FSS plot of $\overline{\ln \Delta}$ for $u=0.01$ ($\circ$), 0.02 ($\bullet$) and 0.03 ($\Box$).}
\label{fig:gap}
\end{figure}
\begin{figure}
\caption{The FSS plot of $\overline{\Delta E_G}/N$ for $u=0.01$ ($\circ$), 0.02 ($\bullet$) and 0.03 ($\Box$) with (a) $a=2.46$ and (b) $a=2$.}
\label{fig:eg}
\end{figure}
\begin{figure}
\caption{The spatial dependence of the string order $\ostr (l; N )$ for $u=0.1$ ($\circ$), 0.2 ($\bullet$) and 0.3 ($\Box$).}
\label{fig:str1}
\end{figure}
\begin{figure}
\caption{The FSS plot of $\overline{\ostr}(N)$ for $u=0.01$ ($\circ$), 0.015 ($\Box$), 0.02 ($\diamond$) and 0.03 ($\triangle$) for (a) $\beta = 0.37$ and (b) $\beta=2$.}
\label{fig:str2}
\end{figure}
\end{document}